\documentclass[pra,nofootinbib,twocolumn,amsmath,amssymb,superscriptaddress]{revtex4-2}
\usepackage{graphicx,amsmath,relsize,epstopdf,color,mathtools,bm,newtxtext,newtxmath,braket,rotating}
\usepackage[hyphenbreaks]{breakurl}
\usepackage[colorlinks=true,linkcolor=blue,citecolor=blue,urlcolor =blue]{hyperref}
\usepackage[normalem]{ulem}

\usepackage{booktabs}
\usepackage[table,xcdraw]{xcolor}

\usepackage{soul,xcolor}

\usepackage{dsfont}

\begin{document}

\title{A double-clad ytterbium-doped tapered fiber with circular birefringence \\ as a gain medium 
for structured light}

\author{Iuliia Zalesskaia}
 \affiliation{Tampere University, Korkeakoulunkatu 3, Tampere, 33720, Finland}
 \email{iuliia.zalesskaia@tuni.fi}

\author{Yuhao Lei}
\affiliation{Optoelectronics Research Centre, University of Southampton, Southampton, SO17 1BJ, United Kingdom
}%

\author{Peter G. Kazansky}
\affiliation{Optoelectronics Research Centre, University of Southampton, Southampton, SO17 1BJ, United Kingdom
}%

\author{Katrin Wondraczek}
\affiliation{Dept. Leibniz Institute of Photonic Technology e.V., Albert-Einstein-Str. 9, 07745 Jena, Germany
}%

\author{Regina Gumenyuk}
 \affiliation{Tampere University, Korkeakoulunkatu 3, Tampere, 33720, Finland}
\affiliation{Ampliconyx Ltd, Lautakatonkatu 18, Tampere, 33580, Finland
}%

\author{Valery Filippov}
\affiliation{Ampliconyx Ltd, Lautakatonkatu 18, Tampere, 33580, Finland
}%

\date{\today}

\begin{abstract}
Amplifying radially and azimuthally polarized beams is a significant challenge due to the instability of the complex beam shape and polarization in inhomogeneous environment. In this Letter, we demonstrated experimentally an efficient approach to directly amplify cylindrical-vector beams with axially symmetric polarization and doughnut-shaped intensity profile in a picosecond MOPA system based on a double-clad ytterbium-doped tapered fiber. To prevent polarization and beam shape distortion during amplification, for the first time to the best of our knowledge, we proposed using the spun architecture of the tapered fiber. In contrast to an isotropic fiber architecture, a spun configuration possessing nearly-circular polarization eigenstates supports stable wavefront propagation. Applying this technique, we amplified the cylindrical-vector beam up to 22 W of average power with 10 ps pulses at a central wavelength of 1030 nm and a repetition rate of 15 MHz, maintaining both mode and polarization stability. 
\end{abstract}

\maketitle

Cylindrical-vector beams with axially symmetric polarization and doughnut-shaped intensity profiles have garnered significant interest in recent decades. This attention is attributed to the broad range of applications, such as high-resolution microscopy \cite{Meng:18, Kozawa:18}, plasmons excitation \cite{DONG2019197}, particle acceleration \cite{app3010070} and trapping \cite{Zhan:04,Dholakia2011}, optical data storage \cite{Zhang:09} and high-performance material processing \cite{Meier2007,Allegre_2012}. However, building a stable, high-power, short-pulse source with high polarization purity and homogeneous annular mode field distribution presents a challenging task due to high mode instability under environmental disturbance. The typical approach for generating high power radially and azimuthally polarized beams involves excitation of the beam by a spatially-variant waveplate (S-waveplate) and then amplification in an isotropic large mode area (LMA) fiber \cite{Lin:14,Lin:17}. This approach has been implemented for both continuous wave (CW) \cite{Lin:14} and nanosecond laser systems \cite{Lin:17}. However, internal inhomogeneity of the isotropic LMA fiber unavoidably leads to the distortion of the spatial intensity and polarization distribution along the propagation and amplification process, limiting the beam quality performance. 
This fiber inhomogeneity originates from the number of external and internal perturbations such as bends, transversal compression, elliptical core, and built-in stresses due to variation of dopants distribution \cite{Fedotov:21}. These perturbations are inevitable characteristics of any isotropic fibers becoming even more pronounced in LMA fibers due to large core area.
Recently, we have demonstrated a promising approach for improvement of LMA fiber homogeneity and decrease of internal stresses by means of imperfection averaging along the fiber length. This can be done by rapid rotation of the preform during the drawing process resulting in high symmetrization of the core and leading to low internal birefringence. This fiber so-called spun tapered double-clad fibers (sT-DCF) \cite{Fedotov:21,Fedotov:212} features nearly circular polarization Eigenstates maintaining the complex polarization of a beam with minimal distortion. In addition, the tapered longitudinal profile of LMA fiber is a proven technique for the elevation of thresholds for nonlinear effects and efficient direct amplification of short pulses up to MW-level peak power and several hundred watts of average power with several tens of microjoules pulse energy \cite{Petrov:2020}. Therefore, the combination of spun and taper architectures has great potential to become a versatile technique for delivering high-power short pulsed signals with complex spatial and polarisation profiles.

In this paper, we demonstrate a successful amplification of cylindrical-vector beams carrying short pulses in Yb-doped sT-DCF retaining both beam structure and polarization profiles. The laser system delivers up to 22 W of average power at the central wavelength of 1030 nm with 10 ps pulses at a repetition rate of 15 MHz. We compare the achieved results in the same laser system with the amplification in isotropic Yb-doped tapered double-clad fiber to confirm the advantage of the spun architecture in maintaining the beam shape and polarization stability.

\begin{figure}[ht]
\centering
\includegraphics[width=\linewidth]{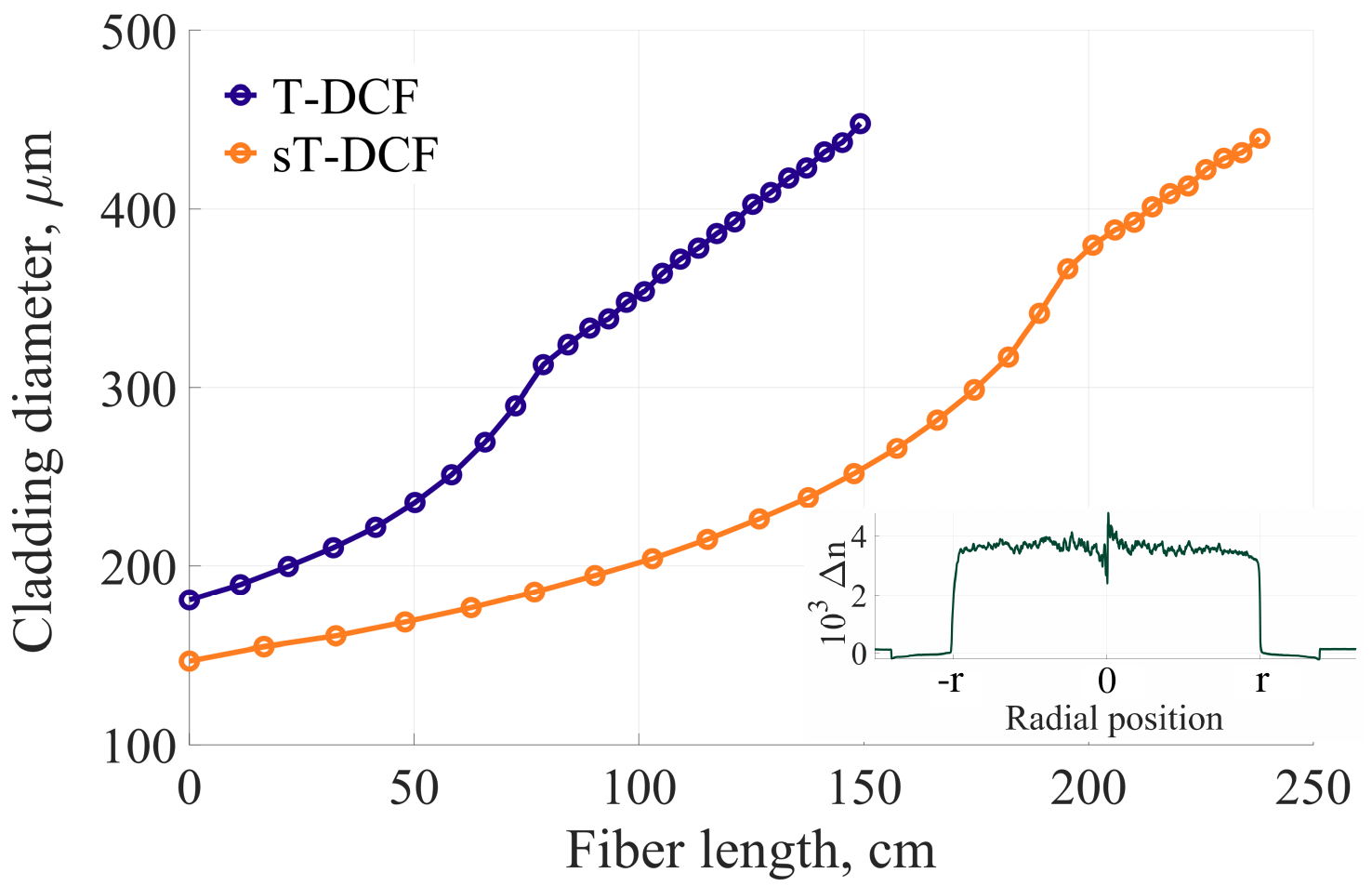}
\caption{T-DCF and sT-DCF longitudinal profiles. The inset shows a cross-section of the refractive index profile of the initial core material.}
\label{fig:profiles}
\end{figure}

Both sT-DCF and tapered double-clad fibers (T-DCF) were manufactured by using a step-index Yb-doped preform produced by REPUSIL technology. A cross-sectional refractive index profile of the initial core material is illustrated in Figure \ref{fig:profiles} (inset). At a wavelength of 976 nm, the in-core absorption coefficient was measured to be 650 dB/m, while the numerical aperture was determined to be 0.1. The preform was constructed with a core-cladding-diameter ratio (CCDR) equal to 11.5. In order to achieve the desired fiber tapering, a continuous variation of the fiber drawing speed was applied. Additionally, for the sT-DCF, a constant preform rotation velocity was implemented during the drawing process to impart the spun architecture. The fibers underwent a tapering process wherein the outer cladding diameter was gradually reduced along a 7 m length, resulting in a final cladding diameter range of 90 µm to 500 µm. To ensure the undistorted propagation of the TM\textsubscript{01} mode, the T-DCF were cut to core/cladding diameters ranging from 16/185 µm to 40/460 µm, while the sT-DCF were cut to a core/cladding diameter ranging from 12.7/146 µm to 37.5/431 µm. The cladding diameter fibers profiles are shown in Figure \ref{fig:profiles}. A V-number of the thin side was equal to 5.04 at 1030 nm for T-DCF and 4 for sT-DCF. 

\begin{figure}[ht]
\centering
\includegraphics[width=\linewidth]{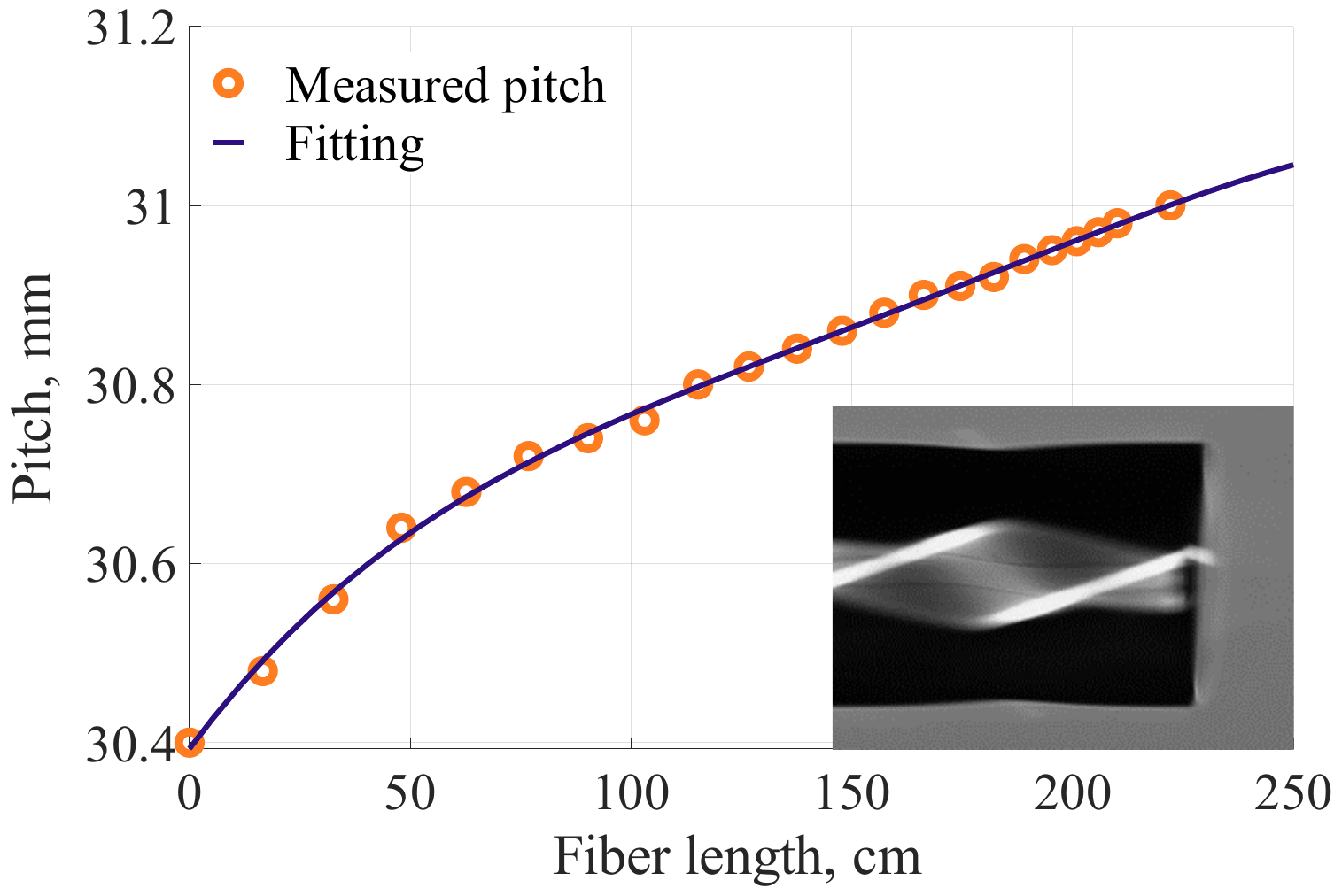}
\caption{Pitch variation of the sT-DCF along the length. The inset shows the side view of the sT-DCF.}
\label{fig:pitch}
\end{figure}

Due to the variation in the drawing speed during the tapering process, the pitch of the sT-DCF slightly changed along the length of the fiber in conjunction with the diameter reduction. The measured pitch sT-DCF profile is presented in Figure \ref{fig:pitch} with an inset of the side view of the sT-DCF. The pitch value of sT-DCF slightly increased from 30.4 mm at the thinnest side to 31 mm when the sT-DCF core diameter reached 37.5 µm.

\begin{figure*}[ht]
\centering
\includegraphics[width=\textwidth]{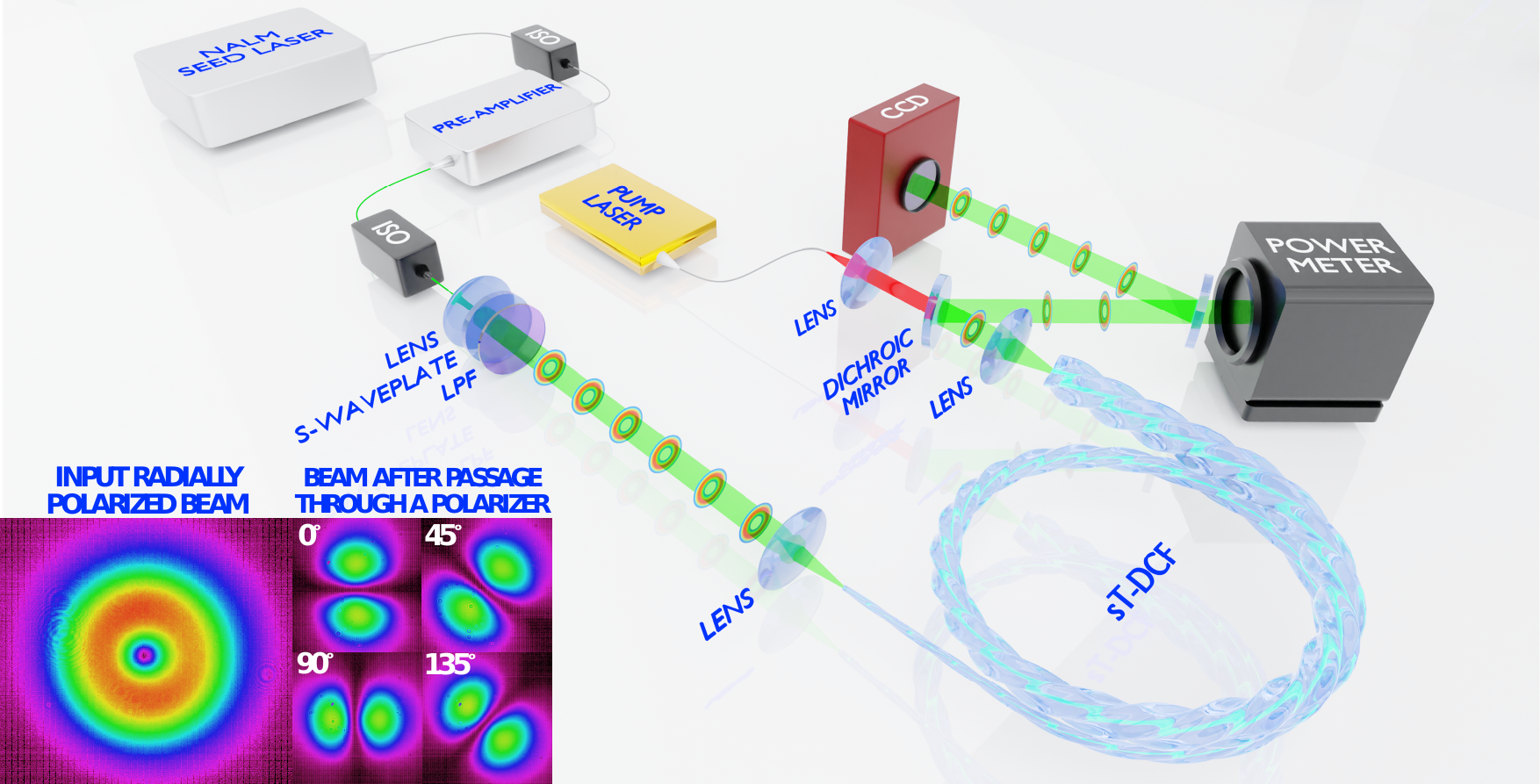}
\caption{The schematic of the experimental setup, ISO - optical isolator, CCD - charged coupled device, LPF - long-pass filter. Inset: a converted by S-waveplate radially polarized beam (left); the radially polarized beam after passage through a rotated linear polarizer at four polarizer axis positions (0\textdegree, 45\textdegree, 90\textdegree, 135\textdegree)  (right).}
\label{fig:setup}
\end{figure*}

In the experimental configuration, a linearly-polarized beam carrying 10 ps pulses at a 15 MHz repetition frequency was generated by a seed laser mode-locked with a nonlinear amplifying loop mirror (NALM) as depicted in Fig. \ref{fig:setup}.

These pulses were subsequently guided into the Yb-doped fiber pre-amplifier, after which they were ultimately yielding an average output power of 50 mW. This stage includes an approximately 160 cm long segment of polarization-maintaining (PM), Panda-type Ytterbium-doped Fiber (YDF) (Coherent PM-YSF-HI-HP) with a core diameter of 6 µm and a cladding diameter of 125 µm. A 976 nm single-mode laser diode supplied 250 mW pump power was used for pumping the PM-YDF, effectively amplifying the pulses. The resulting beam was then directed towards an S-waveplate, where a radially polarized beam was formed. With a diameter of 5 mm, the S-waveplate, was imprinted using birefringence patterning on an infrared-grade synthetic silica glass substrate with an ultrafast laser, operating at 1030 nm with a pulse duration of 600 fs. The waveplate's retardance was tailored to half of the 1050 nm wavelength, and its slow axis azimuth was set at half the polar angle. Writing parameters, including pulse energy, scan speed, and a focusing objective NA of 0.16, were optimized to produce a birefringent X-type modification with a characteristic feature of randomly distributed anisotropic nanopores in silica glass. This specific type of birefringence is characterized by its exceptionally high transmission \cite{Sakakura2020}. The measured transmission of the fabricated S-waveplate exceeded 99\% with reference to an untreated silica glass sample at a wavelength of 1050 nm and the damage threshold was about 1.56 J/cm\textsuperscript{2} for a 1030 nm, 300 fs laser beam \cite{Shayeganrad:22}, which is close to pristine silica glass. The transformed radially polarized beam exhibited a prominent doughnut-shaped intensity profile in the far field, as depicted in the inset Fig. \ref{fig:setup}.

\begin{figure}[ht]
\centering
\includegraphics[width=\linewidth]{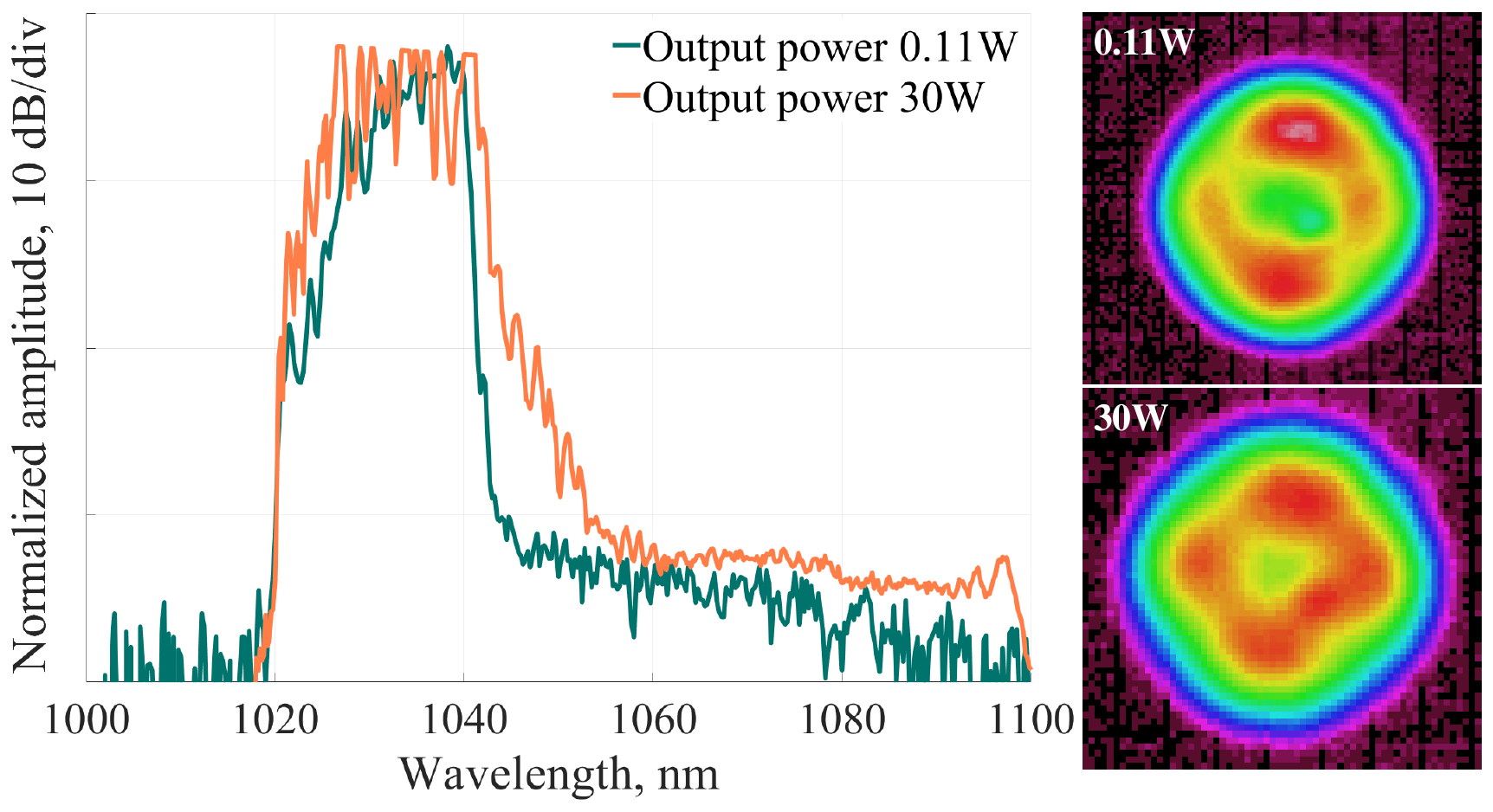}
\caption{The experimental T-DCF output spectrums for 0.11 W and 30 W output power (left). The experimental output intensity distributions in the near-field for 0.11 W (top right) and 30 W  (bottom right) output power.}
\label{fig:tdcfspectrum}
\end{figure}

The beam quality parameter M² for the converted beam was nearly 2.11, which closely corresponds to the theoretical value of 2 for radially polarized beams. The beam was guided through a rotated linear polarizer featuring a two-lobed intensity profile for 0\textdegree, 45\textdegree, 90\textdegree and 135\textdegree polarizer axis (Fig. \ref{fig:setup} (inset)) indicating that the resultant vector beam maintains high polarization distribution uniformity. To safeguard the pre-amplifier from potential damage due to the backward pump of the tapered fiber, a long-pass filter was integrated into the setup. Finally, the radially polarized beam was injected into the high-power amplifier. 
In the first experiment, we employed a T-DCF as our amplification medium. The primary amplifier stage consisted of a T-DCF with a length of approximately 2.5 meters. The thin input side of the T-DCF featured a core diameter of 16 µm and a numerical aperture (NA) of 0.1, while the opposing side had a larger core diameter of about 40 µm. With an input V-number of 5.04 at 1030 nm, the T-DCF is capable of supporting the propagation of the TM\textsubscript{01} mode. To avoid any excess propagation loss, the T-DCF was loosely coiled with a large bend diameter of around 30 cm. The thinner side of T-DCF was perpendicularly cleaved in order to achieve the highest efficiency of radially polarized beam injection without any distortion. Meanwhile, the output end facet was angle-cleaved at an angle of approximately 4 degrees to prevent back reflection. The T-DCF was counter-pumped using a fiber-coupled 976 nm high-power wavelength-stabilized diode laser (BWT K976AG1RN), pigtailed with a 105 µm core diameter fiber and a numerical aperture of 0.22. In this configuration, the final stage of the amplifier reached an optical-to-optical efficiency of 46\% from the total pump power to the T-DCF output power, yielding an output power of 30 W measured directly at the fiber output. 
We conducted an extensive analysis of the beam quality under varying pump power levels. For a relatively low pump power of 5W, which generated an output power of 0.11 W, the measured beam quality was 2.17. When the pump power was increased to 65W, resulting in an output power of 30 W and the beam quality slightly rose to 2.21. The output beam intensity profiles and spectrums corresponding to these measurements can be found in Figure \ref{fig:tdcfspectrum}. 

\begin{figure}[ht]
\centering
\includegraphics[width=\linewidth]{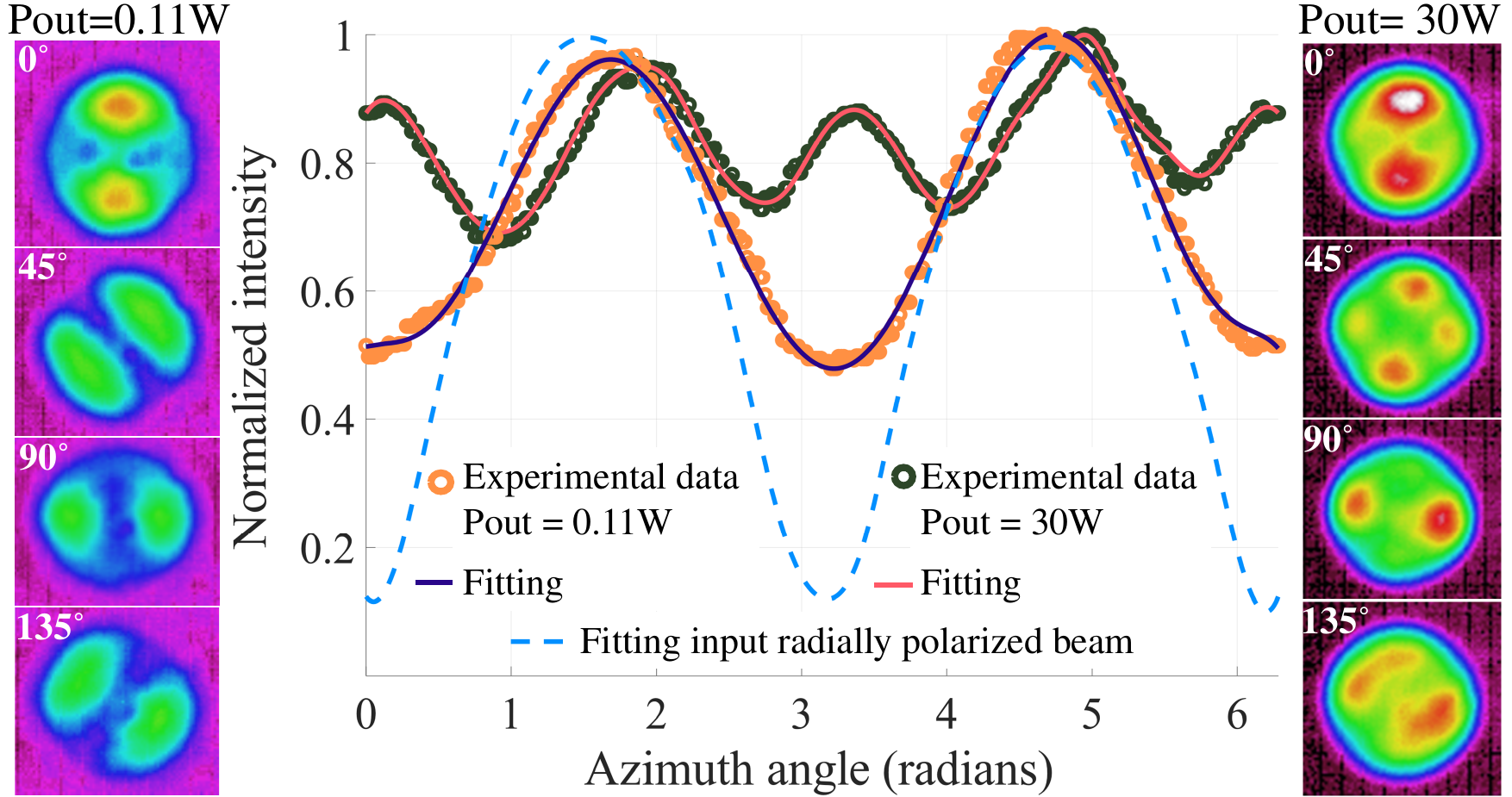}
\caption{The 0.11 W output T-DCF beam after passage through a rotated linear polarizer for four polarizer axis 0\textdegree, 45\textdegree, 90\textdegree, 135\textdegree (left). The normalized azimuthal-intensity profile for radius r after T-DCF beam passage through a linear polarizer (center). The 30 W output sT-DCF beam after passage through a rotated linear polarizer for four polarizer axis 0\textdegree, 45\textdegree, 90\textdegree, 135\textdegree (right).}
\label{fig:polarisationpuritynonspun}
\end{figure}

Upon closely examining the beam intensity profile images, it appears that the beams display some degree of distortion.  To provide further evidence for this observation, we measured the azimuthal-intensity profile for radius r after
beam passage through a linear polarizer. The normalized distribution of these measurements is illustrated in Figure \ref{fig:polarisationpuritynonspun} (center).

To further investigate the polarization properties of the output beams, we introduced a polarizer to the output beam path for both the slightly pumped propagated beam and the highly pumped one. As the intensity profile of the 30 W beam passed through a rotating linear polarizer (depicted in Figure \ref{fig:polarisationpuritynonspun}), for four polarizer axes (0\textdegree, 45\textdegree, 90\textdegree, 135\textdegree) it did not exhibit a pure two-lobes form. This observation suggests that the resulting vector beam may possess a relatively low polarization distribution uniformity.

\begin{figure}[ht]
\centering
\includegraphics[width=\linewidth]{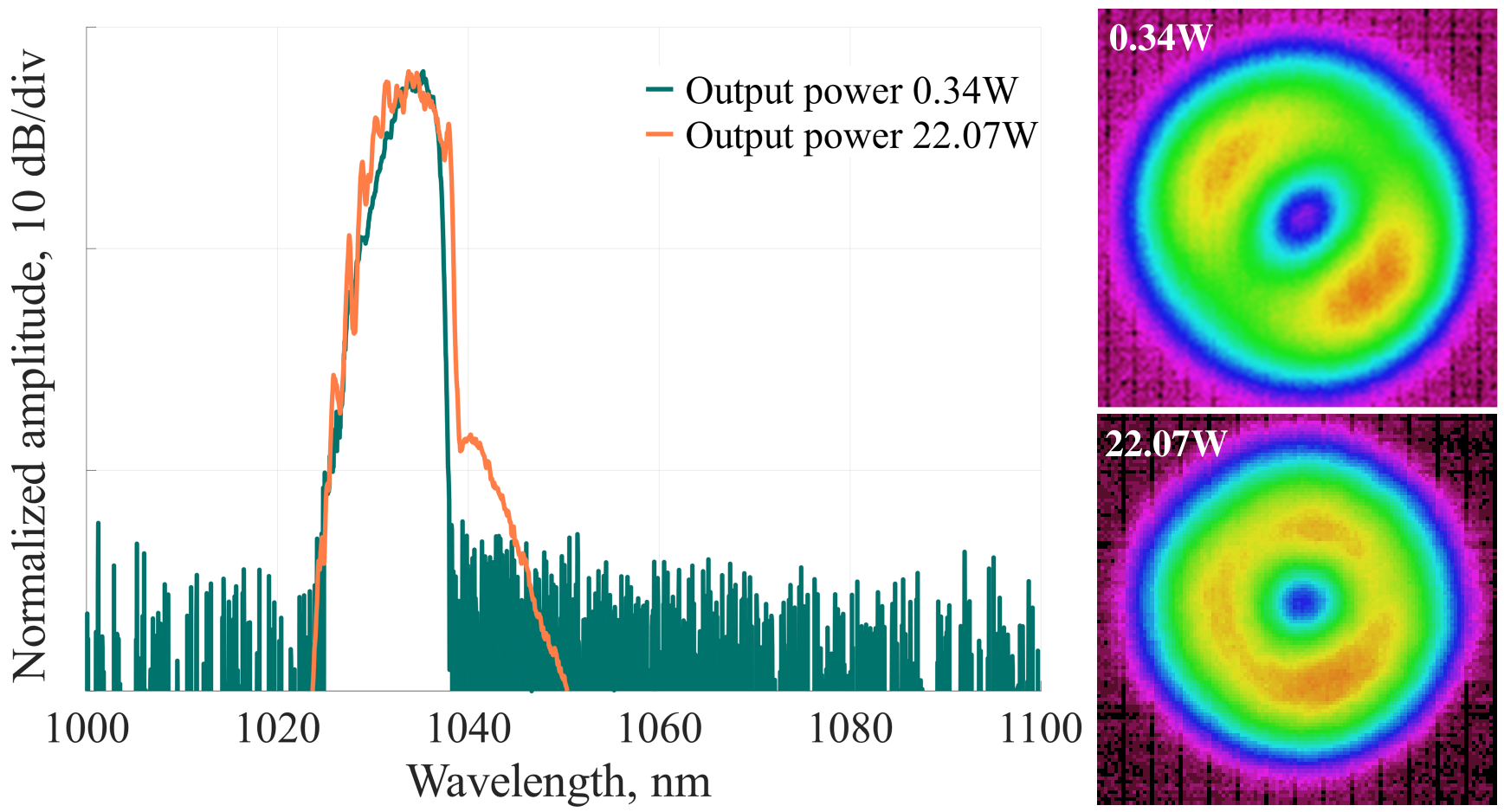}
\caption{The experimental sT-DCF output spectrums for 0.34 W and 22.07 W output power (left). The experimental output intensity distributions in the near-field for 0.34 W (top right) and 22.07 W  (bottom right) output power.}
\label{fig:stdcfspectrum}
\end{figure}

In the second experiment, an sT-DCF was investigated as an alternative to the T-DCF utilized in the first experiment, with core diameters ranging from 12.7 µm to 37.5 µm. This configuration, possessing a V-number of 4 from the thin side of the fiber, facilitated the propagation of the radially polarized TM\textsubscript{01} mode within the core. The output beam was examined at various pump power levels. We indicated that with a 5 W pump power, the output power reached 0.34 W, and the beam quality was measured at 1.98. As the pump power increased to 52 W, the output power reached the level of 22.07 W, corresponding to a 42\% optical-to-optical efficiency, and the beam quality experienced a slight improvement to 1.96. The cause for the beam quality being below the theoretical limit of 2 can be elucidated by the dominance of the LP\textsubscript{11} mode on the x-axis at elevated power levels \cite{Lin:14}. The output beam intensity profiles and spectrum corresponding to these power levels are depicted in Figure \ref{fig:stdcfspectrum}. 

\begin{figure}[t]
\centering
\includegraphics[width=\linewidth]{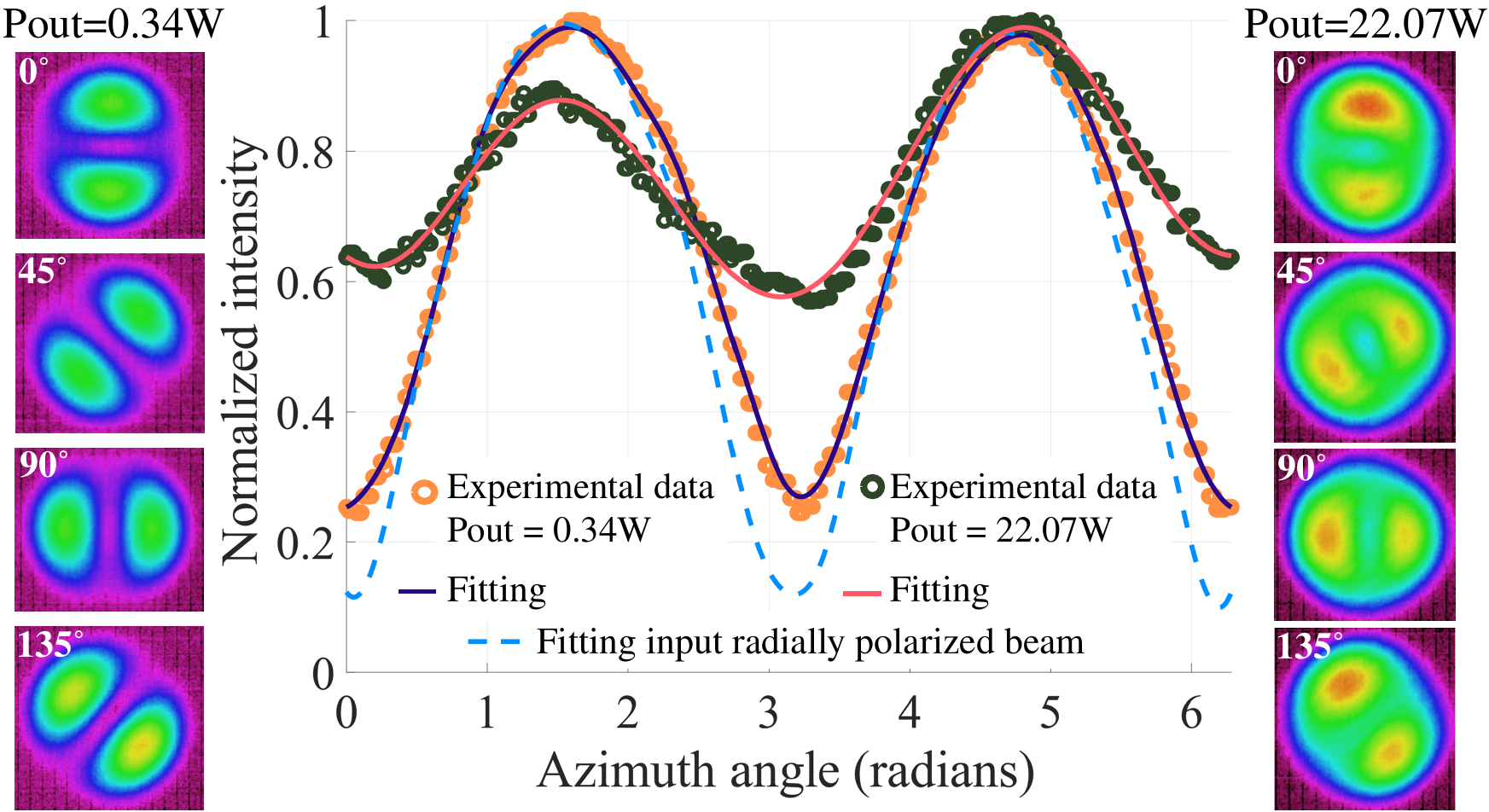}
\caption{The 0.34 W output sT-DCF beam after passage through a rotated linear polarizer for four polarizer axis 0\textdegree, 45\textdegree, 90\textdegree, 135\textdegree (left). The normalized azimuthal-intensity profile for radius r after sT-DCF beam passage through a linear polarizer (center). The 22.07 W output sT-DCF beam after passage through a rotated linear polarizer for four polarizer axis 0\textdegree, 45\textdegree, 90\textdegree, 135\textdegree (right).}
\label{fig:polarisationpurityspun}
\end{figure}

To thoroughly investigate the polarization state, as in the first experiment, a rotating linear polarizer was employed. The intensity profiles of the beam after passing through the rotating linear polarizer at four polarizer axis positions (0\textdegree, 45\textdegree, 90\textdegree, 135\textdegree) for both high and low pump power levels are presented in Figure \ref{fig:polarisationpurityspun}, showcasing the preservation of the complex polarization profile.

Additionally, the normalized azimuthal-intensity profile for radius r after passage through a linear polarizer is depicted in Figure \ref{fig:polarisationpurityspun}, further corroborating the preservation of the complex polarization profile. The exceptional preservation of polarization in the sT-DCF can be attributed to its spun architecture, which leads to an upper isotropic nature and subsequently, low birefringence.
\break

\vspace{100mm} 

In conclusion, we demonstrated the robust solution for the amplification of cylindrical-vector beams with axially symmetric polarization and a doughnut-shaped intensity profile. The active spun double-clad tapered fiber maintained high modal purity during the amplification process due to its special transverse and longitudinal architecture. The radially-polarized beam carrying 10 ps at 10 MHz repetition rate was successfully amplified up to 22 W at 1030 nm retaining both spectral and polarization profiles.

\begin{acknowledgments}
I. Zalesskaia thanks the TLTO doctoral school for the financial support of her PhD study.
\end{acknowledgments}

\textbf{Funding.}  Horizon Europe research and innovation programme (101096317), Academy of Finland (320165), European Research Council (ENIGMA, 789116).

\textbf{Disclosures.} The authors declare no conflicts of interest.

\bibliography{apssamp}

\end{document}